\documentstyle[aps,psfig,epsf,twocolumn]{revtex}
\begin{document} \draft
\title{Maximum-likelihood estimation of the density matrix}
\author{K. Banaszek\thanks{Permanent address: Instytut Fizyki Teoretycznej,
Uniwersytet Warszawski, Ho\.{z}a~69, PL--00--681~Warszawa, Poland}, G. M.
D'Ariano, M. G. A. Paris and M. F. Sacchi}
\address{Theoretical Quantum Optics Group, INFM -- Unit\`a di Pavia \\
Dipartimento di Fisica `A. Volta', Universit\`a di Pavia \\  via
A. Bassi 6, I-27100 Pavia, ITALY}\date{\today} \maketitle
%%%%%%%%%%%%%%%%%%%%%%%%%%%%%%%%%%%%%%%%%%%%%%%%%%%%%%%%%%%%%%%%%%%%%%%%%%%
\begin{abstract}
We present a universal technique for quantum state estimation based on 
the maximum-likelihood method. This approach provides a
positive definite 
estimate for the density matrix from a sequence
of measurements performed on identically prepared copies of the system. The
method is versatile and can be applied to multimode radiation fields
as well as to spin systems. The incorporation of physical
constraints, which is natural in the maximum-likelihood strategy,
leads to a substantial reduction of statistical errors. 
Numerical implementation of the method is based on a particular form
of the Gauss decomposition for positive definite Hermitian matrices.
\end{abstract}


In quantum mechanics, the achievable information on a physical system
is encoded into the density matrix $\hat\varrho$, which allows one to
evaluate all possible expectation values through the Born statistical
rule $\langle\hat O\rangle = \hbox{Tr}(\hat\varrho\hat O)$.  In order
to obtain full information on a quantum system we need to estimate its
density matrix.  In principle, this can be accomplished by successive
measurements on repeated identical preparations of the same system.
With a proper choice of the measurements, and after collecting a
suitably large number of data, we can arrive at reliable knowledge of
the quantum state of the system.

The problem of inferring the complete quantum state from experimental
data has received a lot of attention over past several years. Physical
systems whose quantum state has been fully characterized in recent
experiments, include now a single light mode \cite{QHT}, a diatomic
molecule \cite{Molecule}, a trapped ion \cite{Ion}, and an atomic beam
\cite{Beam}. These fascinating advances stimulate further theoretical
research in two main directions: on one hand, in implementing
effective measurement schemes that connect the density matrix to
directly observable quantities.  On the other hand, in designing
efficient data processing algorithms in a practical experimental setup
in order to extract the optimal amount of information on the quantum
state. In a laboratory, we always deal with finite ensembles of copies
of the measured system \cite{FiniteEnsembles}.  In addition, the
process of detection is usually affected by various
imperfections. This implies the need of developing novel tools
specifically designed to process realistic and finite 
experimental samples.

In this Communication we present a general-purpose method for quantum
state estimation based on the maximum-likelihood (ML) approach
\cite{MaxLik}. We consider statistical treatment of a sample of
measurements performed on repeated preparations of a given system. The
approach presented in this Communication is very general: it allows
one to extract the information on the quantum state from data
collected in a generic scheme, without assuming any specific form of
the measurement.  Its principle of operation is to find the quantum
state that is most likely to generate the observed data. This idea is
quantified and implemented using the concept of the likelihood
functional.

The ML strategy is an entirely different approach to quantum state
measurement compared to the standard quantum-tomographic techniques
\cite{DLP,BanaJMO99}. In quantum tomography the expectation value of
an operator is obtained by averaging a special function (so called
``pattern function'') of experimental data of a sufficiently complete
set of observables---a ``quorum'' of observables. In homodyne
tomography the quorum observables are the quadratures of the
e.m. field for varying phase with respect to the local
oscillator. Hence, typically, a matrix element of the quantum state is
obtained by averaging its pertaining pattern function over data. This
method is very general and efficient, however, in the averaging
procedure, the matrix elements are allowed to fluctuate statistically
through negative values, with resulting large statistical errors.

\par In contrast, the 
ML method estimates the quantum state as a whole. Such a
procedure incorporates {\em a priori} knowledge about relations between
elements of the density matrix. This guarantees positivity 
and normalization of matrix, with the result of a substantial
reduction of statistical errors. These advantages 
of the ML approach are inevitably related to increased
computational complexity of the estimation procedure, which remains a
highly nontrivial problem even if we resort to numerical means.  To the
best of our knowledge, we present in this Communication the first general
solution to this problem, which provides an effective numerical algorithm
for the ML estimation of the density matrix.

We start with the derivation of the likelihood functional ${\cal
L}(\hat{\varrho})$, which links the raw experimental results with the
object to be reconstructed, i.e.\ the density matrix. The physical
situation we have in mind is an experiment consisting of $N$
measurements performed on identically prepared copies of a given
system. Quantum mechanically, each measurement is described by a
positive operator-valued measure (POVM). The outcome of the $i$th
measurement corresponds to the realization of a specific element of
the POVM used in the corresponding run. We shall denote this element
by $\hat{\cal F}_i$.  The likelihood functional ${\cal
L}(\hat{\varrho})$ describes the probability of obtaining the set of
outcomes for a given density matrix $\hat{\varrho}$. For measurements
performed on repeated preparations of the system, it is given by the
product \begin{equation} {\cal L}(\hat{\varrho}) = \prod_{i=1}^{N}
\text{Tr}(\hat{\varrho} \hat{\cal F}_i)\:.  \end{equation} After the
experiment is performed, the operators $\hat{\cal F}_i$ are determined
by the outcomes of the measurements. The unknown element of the above
expression, which we want to infer from our data, is the density
matrix describing the measured ensemble. The general estimation
strategy of the ML technique is to maximize the likelihood functional
over the set of the density matrices.  Several properties of the
likelihood functional are easily found, if we restrict ourselves to
finite dimensional Hilbert spaces.  In this case, it can be easily
proved that ${\cal L}(\hat{\varrho})$ is a concave function defined on
a convex and closed set of density matrices.  Therefore, its maximum
is achieved either on a single isolated point, or on a convex subset
of density matrices. In the latter case, the experimental data are
insufficient to provide a unique estimate for the density matrix using
the ML strategy.  On the other hand, existence of a single maximum
allows us to assign unambiguously the ML estimate for the density
matrix.  This estimate satisfies all the physical constraints, such as
normalization and positivity.

ML estimation of the quantum state, despite its elegant
general formulation, presents a highly nontrivial constrained
optimization problem, even if we resort to purely numerical means. The
central difficulty lies in the appropriate parameterization of the set
of all density matrix. The parameter space should be of the minimum
dimension in order to preserve the maximum of the likelihood function
as a single isolated point. Additionally, the expression of quantum
expectation values in terms of this parameterization should enable
fast evaluation of the likelihood function, as this step is performed
many times in the course of numerical maximization.

Here, we introduce a parameterization of the set of density matrices which
provides an efficient algorithm for maximization of the likelihood function.
We represent the density matrix in the form
\begin{equation}
\label{Eq:rhoTT}
\hat{\varrho} = \hat{T}^{\dagger} \hat{T}\;,
\end{equation}
which automatically guarantees that $\hat{\varrho}$ is positive 
and Hermitian. The remaining condition of unit trace
$\text{Tr}\hat{\varrho} = 1$ will be taken into account using the
method of Lagrange multipliers. In order to achieve the minimal
parameterization, we assume that $\hat{T}$ is a complex lower
triangular matrix, with real elements on the diagonal. This form of
$\hat{T}$ is motivated by the Cholesky decomposition known in
numerical analysis \cite{Cholesky} for arbitrary non negative
Hermitian matrix.  For an $M$-dimensional Hilbert space, the number of
real parameters in the matrix $\hat{T}$ is $M+2M(M-1)/2=M^2$, which
equals the number of independent real parameters for a Hermitian
matrix. This confirms that our parameterization is minimal, up to the
unit trace condition.

In numerical calculations, it is convenient to replace the likelihood
functional by its natural logarithm, which of course does not change the
location of the maximum. Thus the function subjected to numerical
maximization is given by
\begin{equation}
\label{eq:lt}
L(\hat{T}) = \sum_{i=1}^{N} \ln \text{Tr}(\hat{T}^\dagger
\hat{T} \hat{\cal F}_i) - \lambda \text{Tr}(\hat{T}^\dagger
\hat{T})\;,
\label{loglik}
\end{equation}
where $\lambda$ is a Lagrange multiplier accounting for normalization
of $\hat \varrho$ that equals the total number of measurements
$N$ \cite{notal}. 
This formulation of the maximization problem allows one to apply
standard numerical procedures for searching the maximum over the 
$M^2$ real parameters of the matrix $\hat{T}$. The examples
presented below use the downhill simplex method \cite{Ameba}.

Our first example is the application of the ML estimation in quantum
homodyne tomography of a single-mode radiation field \cite{DLP}, which
is so far the most successful method in measuring nonclassical states
of light \cite{QHT,noncl}. The experimental apparatus used in this
technique is the homodyne detector.  The realistic, imperfect homodyne
measurement is described by the positive operator-valued measure
\begin{equation} 
\label{Eq:Hxphi}
\hat{\cal H}(x;
\varphi) = \frac{1}{\sqrt{\pi(1-\eta)}} \exp \left( - \frac{(x-\sqrt{\eta}
\hat{x}_{\varphi})^2}{1-\eta}\right)\:, 
\end{equation}
where $\eta$ is the detector efficiency, and  $\hat{x}_\varphi$ is
the quadrature
operator, depending on the externally adjustable
local oscillator (LO) phase $\varphi$. 

After repeating the measurement $N$ times, we obtain a set of pairs
$(x_i; \varphi_i)$ consisting of the outcome $x_i$ and the LO phase
$\varphi_i$ for the $i$th run, where $i=1,\ldots, N$. The
log-likelihood functional is given by Eq. (\ref{loglik}) with
$\hat{\cal F}_i\equiv\hat{\cal H}(x_i; \varphi_i)$.  Of course, for a
light mode it is necessary to truncate the Hilbert space to a finite
dimensional basis. We shall assume that the highest Fock state has
$M-1$ photons, i.e.\ that the dimension of the truncated Hilbert space
is $M$.  For the expectation $\mbox{Tr}[\hat{T}^{
\dagger}\hat{T}\hat{\cal H}(x;\varphi)]$ it is necessary to use 
an expression which is explicitly positive, in order to  protect the algorithm
against occurrence of small negative numerical arguments of the
logarithm function. A simple derivation yields
\begin{eqnarray}
\lefteqn{
\mbox{Tr}[\hat{T}^{\dagger}\hat{T}\hat{\cal
H}(x; \varphi)] } & & \nonumber \\
& = &  \sum_{k=0}^{M-1}\sum_{j=0}^{k}
\left|\sum_{n=0}^{k-j} 
\langle k | \hat{T} | n+j \rangle B_{n+j,n} \langle n |
x\rangle e^{in\varphi}\right|^2 \;,
\end{eqnarray}
where $B_{n+j,n} = \left[{{n+j} \choose n} \eta^{n}
(1-\eta)^{j}\right]^{1/2} $ and $ \langle n | x \rangle = H_n(x)
\exp(-x^2/2) / \sqrt{2^n n! \pi^{1/2}} $ are eigenstates of the
harmonic oscillator in the position representation---$H_n (x)$ being
the $n$th Hermite polynomial.

\begin{figure}
\begin{center}
\epsfxsize=.47\hsize\leavevmode\epsffile{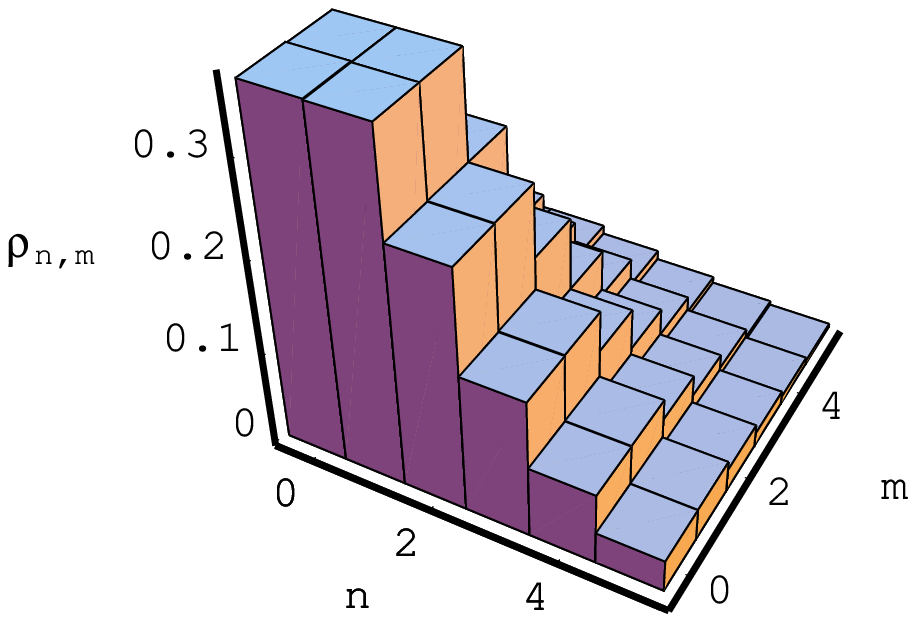}
\hspace{8pt}
\epsfxsize=.47\hsize\leavevmode\epsffile{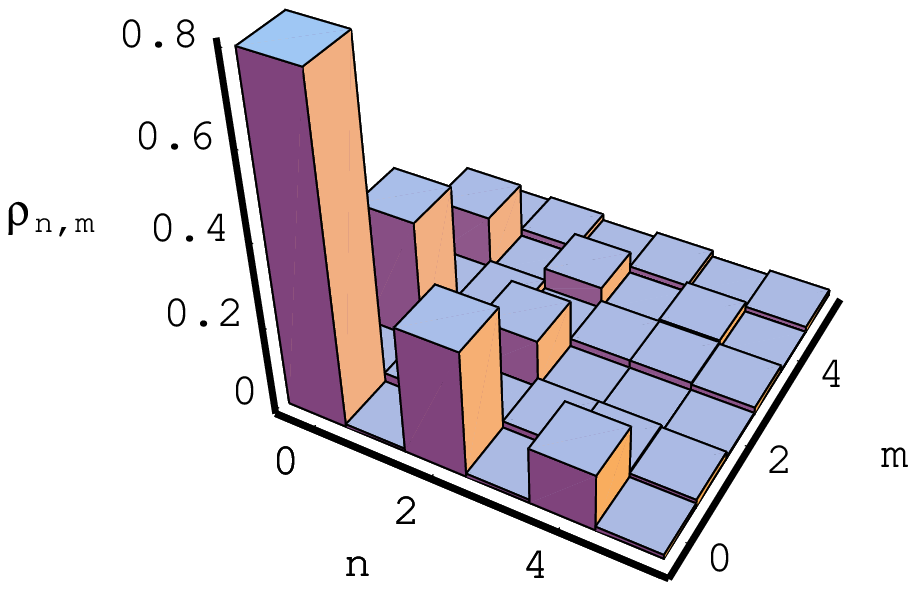}
\vskip .4truecm
\caption{ML reconstruction of the density matrix of a single-mode
radiation field. On the left the matrix elements obtained for a
coherent state with $\langle \hat a^{\dag }\hat a \rangle =1$
photon. On the right for a squeezed vacuum with $\langle \hat a^{\dag
}\hat a \rangle =0.5$ photon. In both cases the ML technique has been
applied to a sample of 50000 simulated homodyne data, and for quantum
efficiency $\eta=80\%$.}\label{Fig:QHT}
\end{center}
\end{figure}

We have applied the ML technique to reconstruct the
density matrix in the Fock basis from Monte Carlo simulated homodyne
statistics.  Fig.~\ref{Fig:QHT} depicts the matrix elements of the
density operator as obtained for a coherent state and a squeezed
vacuum, respectively. Remarkably, only 50000 homodyne data have been
used for quantum
efficiency at photodetectors $\eta=80\%$.

Since statistical aspects of standard quantum homodyne tomography have
been thoroughly studied \cite{QHTErr}, this gives us an opportunity to
compare it with the ML estimation. In the tomographic approach,
statistical errors are known to grow rapidly with decreasing
efficiency $\eta$ of the detector. In contrast, the elements of the
density matrix reconstructed using the ML approach remain bounded, as
the whole matrix must satisfy positivity and normalization
constraints. This results in much smaller statistical errors. As a
comparison one could see that the same precision of the
reconstructions in Fig.~\ref{Fig:QHT} could be achieved using 
$10^7$--$10^8$ data samples with the conventional quantum tomography
of Ref.~\cite{DLP}.  On the other hand, in order to find numerically the ML
estimate we need to set {\em a priori} the cut-off parameter for the
photon number, and its value is limited by increasing computation
time.

\begin{figure}
\begin{center}
\epsfxsize=.47\hsize\leavevmode\epsffile{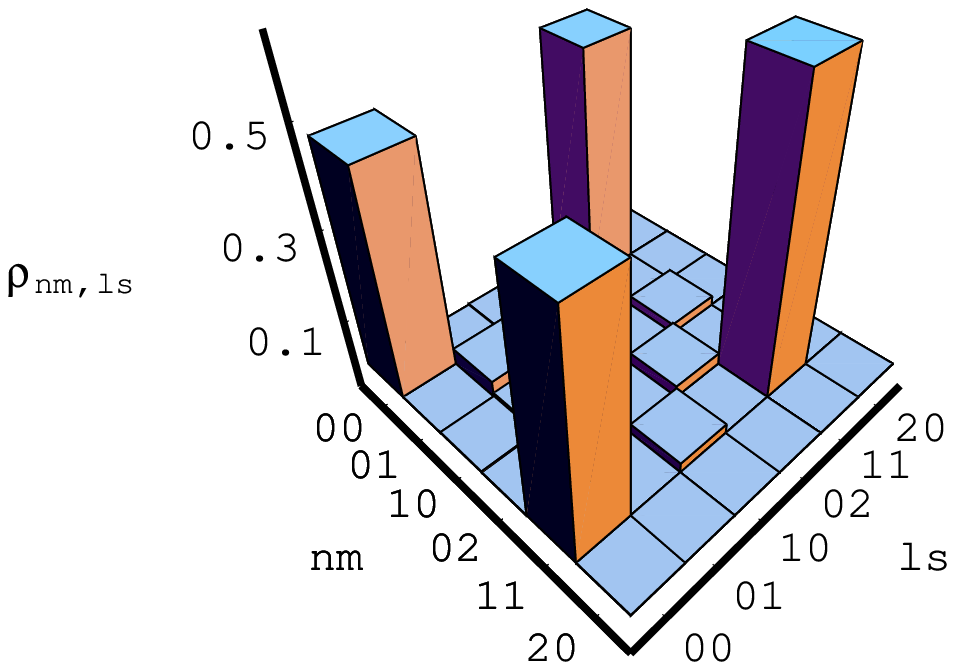}
\hspace{8pt}
\epsfxsize=.47\hsize\leavevmode\epsffile{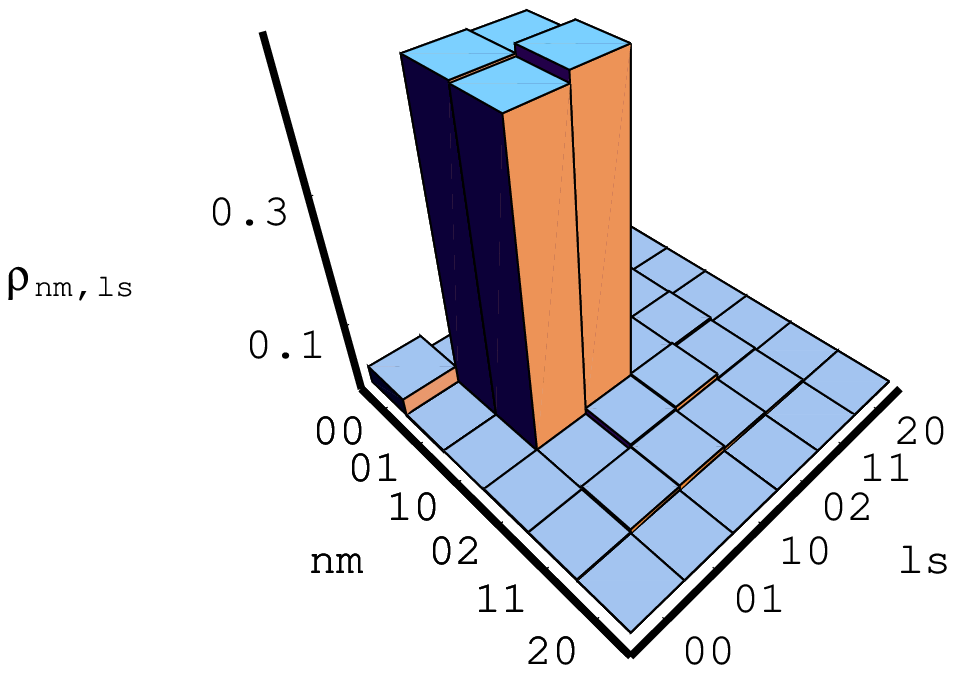}
\vskip .4truecm
\caption{ML reconstruction of the density matrix of a two-mode
radiation field. On the left the matrix elements obtained for the
state $|\Psi _1\rangle =(|00 \rangle + |11 \rangle )/\sqrt 2$; on the
right for $|\Psi _2\rangle =(|01 \rangle + |10 \rangle )/\sqrt 2$. 
For $|\Psi_1 \rangle $ we used
100000 simulated homodyne data and $\eta =80\%$; for $|\Psi_2 \rangle
$ we used 20000 data and $\eta =90\%$.}\label{Fig:TwoMode}
\end{center}
\end{figure}

Another relevant example is the reconstruction of the quantum state of
two-mode field using single-LO homodyning \cite{1lo}.  
Here, the full joint density matrix can be measured by scanning the
quadratures of all possible linear combinations of modes. For two
modes the 
measured quadrature operator is given by $\hat{x}_{\theta\psi_0\psi_1}
= (\hat{a} e^{-i\psi_0}\cos\theta + \hat{b} e^{-i\psi_1} \sin\theta +
\mbox{h.c.} )/\sqrt 2$, where $(\theta,\psi_0,\psi_1)\in S^2 \times
[0,2\pi]$, $S^2$ being the Poincar\'e sphere and one phase ranging
between $0$ and $2\pi$. In each run these parameters are chosen
randomly. The POVM describing
the measurement is given by the right-hand side of
Eq.~(\ref{Eq:Hxphi}), with $\hat{x}_{\varphi}$ replaced by
$\hat{x}_{\theta\psi_0\psi_1}$, and the quantum expectation values of
the POVM can be written as
\begin{eqnarray}
&&\lefteqn{\mbox{Tr}[\hat{T}^{\dagger}\hat{T}\hat{\cal
H}(x; \theta, \psi_0, \psi_1)] 
 = 
\sum_{\shortstack{\scriptsize $k_1, k_2$ \\ \scriptsize $j, n_2$}}
\left|
\sum_{\shortstack{\scriptsize $m_1,m_2$ \\ \scriptsize $n_1$}}
\langle k_1 k_2 | \hat{T} | m_1 m_2 \rangle \right.} \nonumber \\
&& \left. 
\vphantom{\sum_{\shortstack{\scriptsize $m_1,m_2$ \\ \scriptsize $n_1$}}}
\times
\langle m_1 m_2 | \hat{U}^{\dagger}(\theta, \psi_0, \psi_1)
| n_1 + j, n_2 \rangle
B_{n_1+j,n_1} \langle n_1 | x \rangle \right|^{2} \,.\!\label{trt}
\end{eqnarray}
We have simulated an experiment for the two orthogonal states $|\Psi
_1\rangle =(|00 \rangle + |11 \rangle)/\sqrt 2$ and $|\Psi _2\rangle
=(|01 \rangle + |10 \rangle )/\sqrt 2$.  We reconstructed the density
matrix in the two-mode Fock basis using the ML technique.  The results
are depicted in Fig.~\ref{Fig:TwoMode}.

\begin{figure}
\begin{center}
\epsfxsize=.55\hsize\leavevmode\epsffile{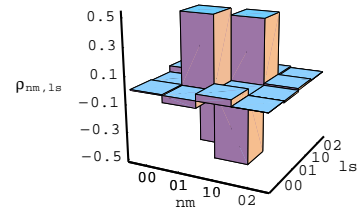}
\vskip .4truecm
\caption{ML reconstruction of the density matrix of a pair of spin-1/2
particles in the singlet state. 
The particles are shared by two parties. In each run, the
parties select randomly and independently from each other a direction
along which they perform spin measurement. 
The matrix elements has been obtained by a sample of 500 simulated
data.}\label{Fig:singlet} 
\end{center}
\end{figure}

Finally, we mention that the ML procedure 
can be applied also for reconstructing the density matrix of spin
systems. For example, let us consider 
$N$ repeated preparations of a pair of spin-1/2
particles. The particles are shared by two parties. In each run, the
parties select randomly and independently from each other a direction
along which they perform spin measurement.
The obtained result is described by the joint
projection operator (spin coherent states)
\begin{equation}
\hat{\cal F}_i = |\Omega^A_{i}, \Omega^B_{i} \rangle \langle
\Omega^A_{i}, \Omega^B_{i} |\;,
\end{equation}
where $\Omega^A_{i}$ and $\Omega^B_{i}$ are the vectors on the Bloch sphere
corresponding to the outcomes of the $i$th run, and the indices
$A$ and $B$ refer to the two particles. As in the previous examples, 
it is convenient to 
use an expression for the quantum expectation value $\mbox{Tr}(\hat{T}^{
\dagger}\hat{T}\hat{\cal F}_i$) which is explicitly positive. 
The suitable form is
\begin{equation}
\mbox{Tr}(\hat{T}^{\dagger} \hat{T}\hat{\cal F}_i)
= \sum_\mu |\langle \mu | \hat{T} | \Omega^A_{i}, \Omega^B_{i} \rangle
|^2\;,
\end{equation}
where $|\mu\rangle$ is an orthonormal basis in the Hilbert space
of the two particles. The result of a simulated experiment with only
500 data for the
reconstruction of the density matrix of the singlet state is shown in
Fig. \ref{Fig:singlet}. 
 
We conclude this Communication with a brief discussion of the
statistical uncertainty of the ML estimate. The
likelihood function can be formally regarded as
a probability distribution on the parameter space. In our case, 
this space is spanned by $M^2$ real parameters which constitute
the triangular matrix $\hat{T}$. We shall denote these parameters
in the vector form as ${\bf t}$. The formal
distribution is given, up to the normalization constant,
by $\delta[\mbox{Tr}(\hat{T}^\dagger \hat{T})-1]\exp L(\hat{T})$.
In the limit of the large number of measurements, $\exp L(\hat{T})$
takes the form of the Gaussian
\cite{Cramer}, with the quadratic form in the exponent
given by the matrix
$
G = - \partial^2 L/\partial {\bf t} \partial {\bf t}'
$.
Furthermore, the constraint $\mbox{Tr}(\hat{T}^\dagger \hat{T}) =1$
means locally orthogonality to the gradient
$
{\bf u} = \partial \mbox{Tr}(\hat{T}^\dagger \hat{T}) /\partial {\bf t}
$.
The covariance matrix for the parameters ${\bf t}$ is consequently
given by \cite{Eadie}
\begin{equation}
V = G^{-1} - \frac{G^{-1} {\bf u} {\bf u}^{T} G^{-1}}{{\bf u}^{T}
G^{-1} {\bf u}}\;.
\end{equation}
With this result, we can estimate errors for the density matrix using
simply the propagation law applied to Eq.~(\ref{Eq:rhoTT}).

Summarizing, we have developed a universal maximum likelihood
algorithm for estimating the density matrix. 
With respect to conventional quantum tomography this method has the
great advantage of needing much smaller experimental samples, making
experiments with low data rates now feasible, however with a 
truncation of the Hilbert space dimension. We have shown that the
method is general and the algorithm has solid
methodological background, its reliability being confirmed in 
a number of Monte Carlo simulations. 

{\em Acknowledgements.} 
We would like to thank Zdenek Hradil for interesting discussions.  This work
has been cosponsored by MURST under the project ``Amplificazione e rivelazione
di radiazione quantistica''. K. B. is supported by INFM
and by KBN Grant 2P03B~089~16.  M. G. A. P. and M. F.
S. are supported by INFM through the project PRA-CAT 1997. 

%%%%%%%%%%%%%%%%%%%%%%%%%%%%%%%%%%%%%%%%%%%%%%%%%%%%%%%%%%

%%%%%%%%%%%%%%%%%%%%%%%%%%%%%%%%%%%%%%%%%%%%%%%%%%%%%%%%%%

\begin{thebibliography}{99}
\bibitem{notal} 
Writing $\hat{\varrho}$ in terms of its eigenvectors 
$| \psi_\mu 
\rangle $ as $
\hat{\varrho} = \sum_{\mu } y_{\mu}^{2} | \psi_\mu 
\rangle \langle \psi_\mu |
$, with real $y_{\mu}$, the maximum likelihood condition 
$\partial L/\partial y_{\nu} = 0$ reads 
\begin{equation}
\lambda y_{\nu} = \sum_{i=1}^{N} \frac{y_\nu \langle \psi_\nu
| \hat{\cal F}_i | \psi_\nu \rangle}
{\hbox{Tr}(\hat\varrho \hat{\cal F}_i)}\;,
\end{equation}
which, after multiplication by $y_{\nu}$ and summation over $\nu $, 
yields $\lambda = N$.
\end{thebibliography}
\end{document}